\documentclass[journal]{IEEEtran}
\usepackage{amsmath,amssymb}
\usepackage{amsfonts}
\usepackage{dsfont}
\usepackage{comment}
\usepackage{algorithmic}
\usepackage{comment}
\usepackage{caption}
\usepackage{multirow}
\usepackage[noabbrev]{cleveref}
\usepackage{balance}
\usepackage{algorithm}
\usepackage{float}
\usepackage{color}
\usepackage{pgfplots}           
\pgfplotsset{compat=1.16}

\usepackage{enumerate}
\usepackage{graphics}
\usepackage{subcaption}
\usepackage{cite}
\usepackage{mathrsfs}
\usepackage{stfloats}

\usepackage{tikz}
\usetikzlibrary{positioning}
\usetikzlibrary{shapes.geometric}
\usetikzlibrary{shapes.misc}
\usepackage{mathdots}
\usepackage{yhmath}
\usepackage{cancel}
\usepackage{color}
\usepackage{siunitx}
\usepackage{array}
\usepackage{multirow}
\usepackage{textcomp}
\usepackage{gensymb}
\usepackage{tabularx}
\usepackage{booktabs}
\usetikzlibrary{fadings}
\usetikzlibrary{patterns}
\usetikzlibrary{shadows.blur}
\usepackage{graphicx}
\usetikzlibrary{decorations.pathreplacing,calligraphy}
\usetikzlibrary{arrows.meta, decorations.markings}
\usepackage{tikz}
\usepackage{tkz-euclide}
\usepackage{pgfplots}

\usepackage{enumitem}

\newtheorem{proposition}{Proposition}
\newtheorem{remark}{Remark}

\newcommand{\dmin}{d_{\mathrm{min}}}

\usepackage{amsmath,amssymb}
\usepackage{amsfonts}

\usepackage{color}
\usepackage{pgfplots}           
\pgfplotsset{compat=1.16}
\usepackage{caption}

\usepackage{algorithm}
\usepackage{algorithmic}

\usepackage{enumerate}
\usepackage{graphics}
\usepackage{cite}
\usepackage{mathrsfs}
\usepackage{subcaption}
\usepackage{stfloats}
\usepackage{xfrac}

\usepackage{mathtools, nccmath}

\makeatother

\usepackage{lastpage}
\begin{document}
	
\title{Elliptic Curve Modulation (ECM) for Extremely Robust Physical Layer Encryption}

\author{
Thrassos K. Oikonomou,~\IEEEmembership{Graduate Student Member,~IEEE,}
and George K. Karagiannidis,~\IEEEmembership{Fellow,~IEEE}
\thanks{The authors are with the Wireless Communications and Information Processing (WCIP) Group, Electrical \& Computer Engineering Dept., Aristotle University of Thessaloniki, 54 124, Thessaloniki, Greece (e-mails: \{toikonom, geokarag\}@ece.auth.gr).}
}
\maketitle	

\begin{abstract} 
This paper introduces Elliptic Curve Modulation (ECM), a novel modulation scheme that can be leveraged to effectively shuffle transmitted data while maintaining symbol error probability (SEP) performance equivalent to unencrypted systems. By utilizing the well-distributed elliptic curve points over the field of large primes, ECM enhances symbol obfuscation, making it a powerful foundation for physical-layer encryption (PLE). Each symbol is mapped from a predefined key while preserving a minimum Euclidean distance constraint, ensuring strong security against adversarial inference without compromising error performance. Building on ECM’s strong obfuscation capabilities, we propose ECM with dynamic rotation (ECM-DR) as a practical PLE scheme that achieves near-maximal obfuscation while balancing precomputation complexity. By leveraging a reduced subset of precomputed elliptic curve points and key-based dynamic constellation rotation, ECM-DR ensures that each transmission remains unpredictable, significantly enhancing security compared to traditional PLE schemes without additional computational cost. Security analysis confirms ECM’s resilience to brute-force attacks, while numerical results demonstrate its strong obfuscation capabilities. Furthermore, ECM-DR achieves near-maximum information entropy while preserving the SEP performance of unencrypted quadrature amplitude modulation (QAM), offering an extremely robust solution for secure wireless communications.
\end{abstract}

\maketitle
\vspace{-5mm}
\section{Introduction}
Wireless communication systems have revolutionized the way we connect and exchange information without physical connections, enabling seamless global communication and supporting critical infrastructures across various industries \cite{6G,SuperConstellations}. As these networks continue to expand, their open nature introduces significant security challenges, as adversaries can exploit emitted signals to conduct traffic analysis, monitor transmission patterns, and extract valuable intelligence about network operations. To counter these threats, research has predominantly focused on upper-layer encryption techniques, which secure data content by ensuring message confidentiality and authentication. However, while these measures provide strong protection, they leave critical physical layer attributes, such as modulation schemes and signal characteristics, exposed, allowing adversaries to passively observe and analyze transmitted signals \cite{pls1, forensics1, forensics2}. Specifically, since these attributes define how information is encoded and propagated over the wireless channel, an eavesdropper can exploit their predictable structure to infer transmission behaviors, monitor network activity, and distinguish between different users or devices. Moreover, this exposure further enables attackers to manipulate network protocols, exploit MAC addresses to bypass access control mechanisms, or launch denial-of-service attacks \cite{pls2}, \cite{pls3}. Thus, rather than solely protecting data content, securing the underlying signal properties becomes essential for maintaining resilience against interception and active exploitation.

To effectively obscure transmitted information, security mechanisms must manipulate the degrees of freedom inherent in wireless signals. 
One fundamental degree of freedom is the signal-to-noise ratio (SNR), which dictates how well a receiver can reconstruct the transmitted signal based on the strength of the received waveform relative to background noise \cite{pls1, pls2, thanos, apostolos, pls3, pls4}. 
Within this context, covert communication represents a well-established physical-layer security technique that utilizes low-SNR signaling to render transmissions statistically undetectable. By embedding signals beneath the noise floor or emulating the statistical properties of the ambient channel, its primary objective is to conceal the existence of communication itself, rather than to protect the transmitted content \cite{covert1}.
However, since a higher SNR improves signal clarity, an eavesdropper can also actively enhance their reception by employing multiple antennas, counteracting any security measures relying solely on deteriorating channel conditions. In this direction, Physical Layer Encryption (PLE) offers a more robust approach by obfuscating transmitted symbols, ensuring that even if an eavesdropper captures a high-quality signal, they are unable to extract any meaningful information \cite{ple1, ple2, ple3, ple4, ple5, ple6}. In more detail, rather than attempting to restrict access to the wireless channel, PLE transforms the structure of the transmitted data itself, making it resistant to analysis regardless of channel conditions. As a result, even if an eavesdropper enhances their SNR using multiple antennas or advanced signal processing techniques, they cannot recover the original information without additional intrinsic knowledge. Moreover, because PLE embeds security directly into the waveform, it remains effective even if upper-layer cryptographic defenses are compromised, ensuring that wireless communications do not solely depend on traditional encryption methods. To this end, PLE establishes a necessary foundation for safeguarding modern wireless networks against eavesdropping and interception. 

\vspace{-3mm}
\subsection{State-of-the-Art}
Recent research in physical-layer encryption (PLE) has introduced a variety of techniques aimed at securing wireless transmissions by modifying the waveform to obscure the structure of the underlying data. Specifically, in \cite{intrinsic-interference1}, the authors proposed injecting imaginary symbols alongside real ones to confuse unauthorized receivers, while \cite{intrinsic-interference2} extended this idea by blending real and imaginary components to enhance obfuscation, though both approaches increased symbol error probability and decoding complexity. To mitigate such effects, structured noise was embedded directly into the waveform in \cite{noise-insertion}, \cite{noise-insertion2}, and \cite{noise-insertion3}, which improved secrecy but required dynamic power allocation that reduced spectral efficiency. This principle was further extended to multicarrier systems in \cite{trung-ofdm}, where dummy subcarriers were used to mask data, although the resulting redundancy reduced bandwidth utilization. In pursuit of greater symbol-level confusion, the authors in \cite{cryptographic-primitives} and \cite{APLE} introduced security matrices that disperse input bits across multiple output symbols through structured mappings, yet these techniques impose high computational complexity and latency. To address processing overhead, a lightweight alternative was proposed in \cite{ple_new}, where power-level modulation was employed to reduce complexity and provide authentication, although it introduced signal distortions that compromised communication reliability. Despite their differences, all of these methods alter the transmitted signal to hinder eavesdropping, which often leads to degraded communication performance, including reduced spectral efficiency, increased error rates, and excessive processing demands. This trade-off reflects a structural limitation of current approaches, which treat encryption as an auxiliary layer rather than a native component of the modulation process. To this end, as modern wireless systems demand both high security and uncompromised performance, it becomes evident that future PLE solutions must embed encryption directly into the modulation framework to ensure robust secrecy without sacrificing communication quality.

Building on this premise, several techniques have emerged that embed encryption directly into the modulation process to enhance physical-layer security without introducing redundancy. Among these, the works in \cite{dynamic-constellation-rotation, Jiang, A-Secure-Phase-Encrypted-IEEE} apply key-based pseudorandom or chaos-driven phase rotations to modulated symbols, allowing the preservation of signal-to-noise ratio and maintaining symbol error probability (SEP). While suitable for systems with strict Quality of Service (QoS) requirements, these methods provide only partial obfuscation, as the amplitude remains unaltered and can be statistically analyzed by an eavesdropper over time. In addition, their reliance on precise phase synchronization introduces implementation complexity, increases latency, and reduces robustness against phase noise. 
To address some of these shortcomings, the symbol-based one-time pad (SOTP) in \cite{sotp} leverages the reciprocity of Rayleigh fading channels to generate symbol-level keys derived from quantized channel state information (CSI), enabling encryption without distorting the transmitted waveform and thereby preserving QoS. However, the resulting symbol distribution follows a Gaussian-like shape that underutilizes the available entropy, and improving security through finer CSI quantization imposes tight synchronization requirements that degrade reliability and exacerbate SEP performance in practical settings. Taken together, these modulation-aware methods represent meaningful progress toward secure waveform design, but their limitations in amplitude obfuscation, entropy maximization, and synchronization tolerance hinder their ability to deliver the symbol unpredictability and operational robustness essential for scalable and future-proof PLE. This observation emphasizes the need for a fundamentally new approach capable of generating highly unpredictable constellations, preserving decoding performance, and avoiding dependency on time-varying channel conditions or complex online operations.

\subsection{Motivation \& Contribution}

To address the limitations highlighted in recent PLE methods, it becomes essential to establish a fundamentally new framework capable of generating highly unpredictable constellations, preserving decoding reliability, and avoiding dependence on time-varying channel conditions or complex online operations. Realizing such a framework requires a mathematically grounded approach that enables structured symbol mappings with high entropy, low complexity, and consistent performance. In this context, elliptic curves offer a particularly powerful foundation, as their rich algebraic structure supports nonlinear transformations that are inherently difficult to reverse, yet analytically tractable. These properties have made elliptic curves central to modern cryptography, where they are widely used in public key generation, digital signatures, and secure key exchange due to the elliptic curve discrete logarithm problem (ECDLP) \cite{elliptic-curve-book}. Additionally, their structured point sets have found application in algebraic coding theory, enabling the design of error-correcting codes with favorable minimum distance characteristics. As a result, these established advantages strongly suggest that elliptic curves can support symbol-level encryption schemes that combine unpredictability with mathematical rigor. To the best of the authors’ knowledge, however, this potential has not yet been leveraged in modulation-based PLE, marking a significant and unexplored opportunity for secure and efficient signal design for enhanced security.

In this paper, we propose Elliptic Curve Modulation (ECM), a novel PLE framework that leverages the structured yet unpredictable distribution of elliptic curve points over discrete fields to achieve secure waveform design. Unlike conventional obfuscation methods, ECM constructs symbol constellations based on predefined keys while enforcing a minimum distance constraint, thereby maximizing information entropy without degrading symbol error performance. 
To support practical deployment, we further develop ECM with Dynamic Rotation (ECM-DR), which applies pseudorandom rotations over a reduced bank of precomputed tuples to balance security and offline complexity. ECM outputs complex-valued symbols and preserves modulation compatibility with standard communication protocols, enabling seamless integration into existing wireless systems such as LTE, 5G NR, and Wi-Fi. 
The key contributions of this work are summarized as follows:
\begin{itemize}
\item We introduce ECM, a novel modulation-based encryption scheme that uses elliptic curve structures to generate high-entropy constellations under minimum distance constraints, ensuring strong symbol obfuscation while maintaining SEP and spectral efficiency.
\item We propose ECM-DR, a scalable variant that achieves near-maximum obfuscation by applying dynamic rotation to a reduced subset of symbols, significantly reducing offline complexity while enabling real-time operation.
\item We present a detailed security analysis, showing that ECM resists brute-force attacks through its structured yet unpredictable symbol mappings and operates independently of channel characteristics or SNR assumptions.
\item We demonstrate that ECM is resilient to quantum threats, as its encryption strength does not rely on assumptions vulnerable to quantum algorithms, and can be paired with post-quantum key exchange mechanisms such as QKD.
\item We provide extensive numerical evaluations showing that ECM-DR achieves near-optimal information entropy and consistently outperforms state-of-the-art PLE methods while preserving SEP parity with unencrypted QAM schemes.
\end{itemize}

\vspace{-4mm}
\subsection{Structure}
The structure of this paper is as follows. Section II introduces the system model along with the necessary preliminaries on elliptic curves, while Section III presents the proposed ECM framework and its design methodology. Section IV describes the integration of ECM into a baseline PLE scheme, and Section V provides the corresponding security analysis. Section VI introduces an enhanced version of the scheme that incorporates dynamic constellation rotation to improve encryption strength and reduce complexity, and Section VII presents the simulation results that validate the proposed methods. Finally, Section VIII concludes the paper.

\section{System Model}

\begin{figure} 
	\centering
	\begin{subfigure}[b]{0.45\textwidth}
		\centering
		\includegraphics[width=\textwidth]{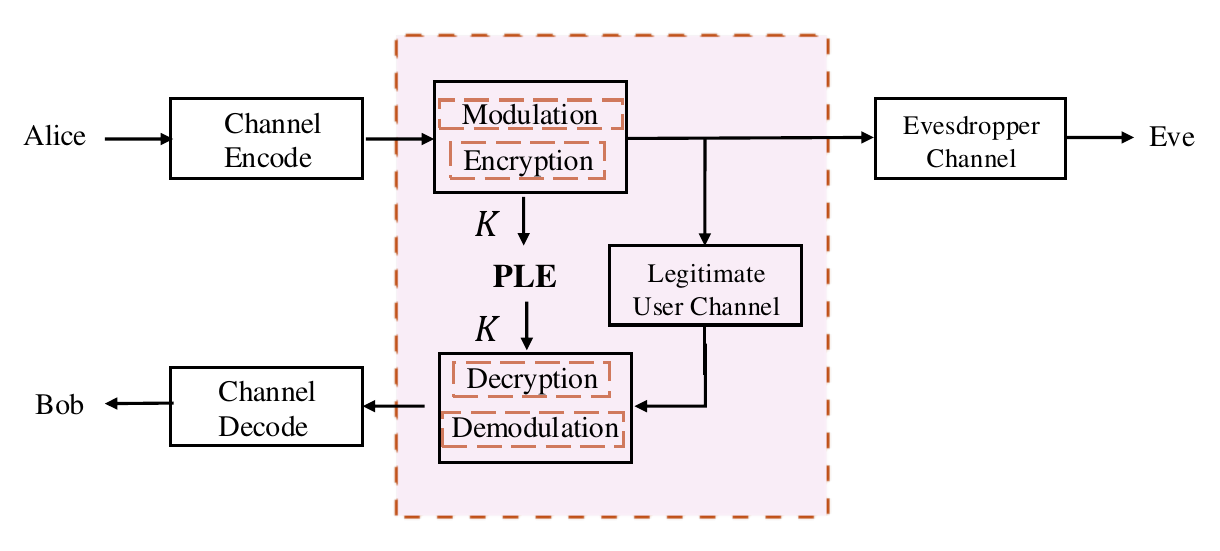} 
	\end{subfigure}
	\hfill 
	\begin{subfigure}[b]{0.45\textwidth} 
		\centering
		\includegraphics[width=\textwidth]{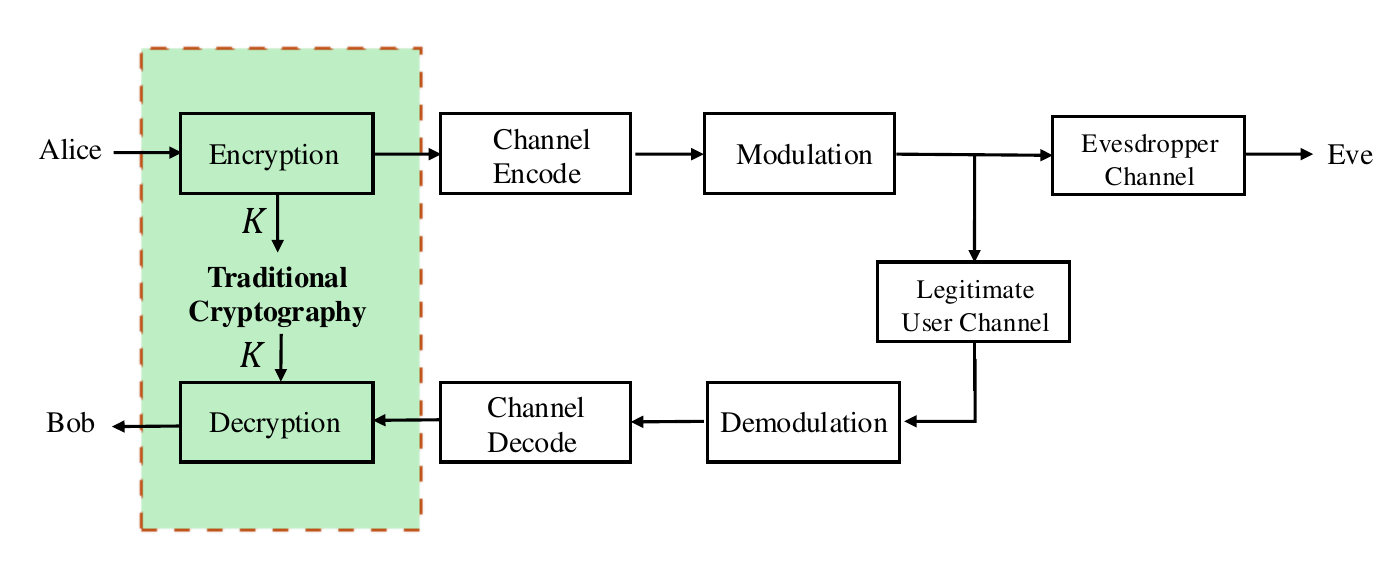} 
	\end{subfigure}
	\caption{PLE versus Traditional Cryptography}
	\label{fig:ple}
\end{figure}

We consider a wireless communication system consisting of a transmitter (Alice),  a legitimate receiver (Bob), and an eavesdropper (Eve), where a PLE scheme is applied at the modulation level. Unlike conventional cryptographic schemes, which operate at higher layers before modulation, PLE integrates encryption directly into the modulation process, as illustrated in Fig. \ref{fig:ple}. 
To establish secure communication, Alice and Bob share a secret key $K$, which can be generated using various protocols such as channel reciprocity, elliptic curve cryptography (ECC), or quantum key distribution (QKD) \cite{channel_reciprocity}. In time-division duplex (TDD) systems, channel reciprocity naturally supports symmetric key generation, while in frequency-division duplex (FDD) systems, where reciprocity does not apply, secure key exchange can still be achieved through cryptographic or quantum-safe alternatives. This flexibility ensures that the ECM framework remains compatible with a wide range of deployment scenarios and evolving security requirements. 
These keys dynamically modify the constellation patterns, ensuring that the modulation symbols are encrypted in real-time. As shown in Fig. \ref{fig:ple}, traditional cryptographic systems leave modulation symbols exposed, making them susceptible to interception and offline analysis. In contrast, PLE alters the transmitted symbols at the physical layer, ensuring that only authorized receivers with the correct keys can reconstruct the original data. Moreover, to model a worst-case scenario, Eve is assumed to have advanced signal processing capabilities and to receive the transmitted signal with an equivalent SNR as Bob, allowing her to observe the channel without degradation. However, since encryption is embedded within the modulation process, an unauthorized receiver cannot extract meaningful information, even with perfect reception, unless the correct keys are applied.

Since the encryption process is embedded at the modulation level, the transmitted signal remains dependent on the dynamically adjusted constellation patterns dictated by the shared secret keys. Specifically, as illustrated in Fig. \ref{fig:ple}, this integration ensures that even if Eve receives the signal under identical conditions as Bob, she cannot decrypt the information without access to the secret keys, as the constellation remains obfuscated. In contrast, as shown in Fig. \ref{fig:ple}, traditional cryptographic systems apply encryption at higher layers, leaving the modulation symbols exposed during transmission. This exposure allows a high-SNR eavesdropper to intercept the raw symbols and process them offline, increasing the risk of unauthorized access despite the presence of upper-layer encryption.

\vspace{-2mm}
\section{Elliptic Curve Modulation (ECM)}

In this section, we introduce ECM, a novel modulation scheme that leverages the geometric and algebraic properties of elliptic curves to construct secure and structured constellations. Specifically, this section provides an overview of elliptic curve fundamentals, establishing the necessary mathematical background for ECM. Then, we formally define ECM, detailing the process of symbol mapping, constellation formation, and power normalization to ensure practical and efficient implementation within wireless communication systems.

\subsection{Elliptic Curve Preliminaries}
Elliptic curves are algebraic structures defined over finite fields that support well-defined group operations and are widely utilized in secure communication and coding due to their ability to perform structured yet unpredictable mappings. In more detail, an elliptic curve $E$ defined over a finite field $\mathbb{F}{p}$, where $p$ is a prime number, is typically expressed by the Weierstrass equation as
\begin{equation} \label{ec_definition}
\begin{aligned}
E: y^2 \equiv x^3 + ax + b \mod{p},
\end{aligned}
\end{equation}
where $a$ and $b$ are elements of $\mathbb{F}{p}$ and are chosen so that the discriminant $\left(4a^3 + 27b^2 \mod{p}\right) \neq 0$. This condition ensures that the curve is non-singular, having no cusps or self-intersections, which is essential for defining a well-structured abelian group under the elliptic curve point addition operation. This group structure, along with scalar multiplication, enables the construction of deterministic symbol mappings that are both efficient and resistant to inference, forming the foundation of the proposed modulation framework.

Within this group, the set of points $(x, y)$ that satisfy the elliptic curve equation, along with a special point at infinity denoted as $\mathcal{O}$, form an abelian group. The fundamental group operation, known as point addition, enables the combination of two points $P$ and $Q$ on the curve to produce a third point that also lies on the curve. Geometrically, this is achieved by drawing a line through $P$ and $Q$, determining its intersection with the curve, and reflecting this intersection point over the $x$-axis. This operation satisfies the essential group properties, ensuring closure, associativity, an identity element (the point at infinity $\mathcal{O}$), and the existence of inverses, thereby maintaining a well-defined algebraic structure.

Building upon point addition, scalar multiplication extends this operation by iteratively adding a point to itself. This operation is fundamental in elliptic curve systems, as it allows transformations that are easy to compute but difficult to reverse. Given an integer $k$, scalar multiplication is defined as
\begin{equation}
    k P = \underbrace{P + P + \dots + P}_{k}.
\end{equation}
The smallest positive integer $n$ for which $n G = \mathcal{O}$ defines the order of a point $G$, characterizing the cyclic nature of the elliptic curve group. To efficiently compute scalar multiplication, algorithms such as the double-and-add method \cite{double-and-add} are commonly employed, enabling fast operations even for large values of $k$.

The difficulty of reversing scalar multiplication forms the basis of the ECDLP, which underpins the security of elliptic curve cryptographic applications \cite{elliptic-curve-book}. Given a generator point $G$ and a resulting point $P = k G$, determining the scalar $k$ is considered computationally infeasible, particularly when $G$ is of large prime order $n$. This security is reinforced by the distribution of points across the finite field, where the resulting set of points exhibits a pseudo-random scattering. As $p$ and $n$ increase, this distribution becomes more uniform, eliminating recognizable patterns that could aid an attacker in reversing the operation \cite{elliptic-curve-book}. By maintaining a large key space and an unpredictable structure, ECDLP remains resistant to brute-force and algebraic attacks, forming the foundation of elliptic curve-based security. These properties also make elliptic curves highly suitable for structured yet unpredictable transformations, a characteristic that can be leveraged in the design of ECM for secure symbol mapping.

\subsection{Elliptic Curve Modulation (ECM)}
The ability of elliptic curves to define a structured yet unpredictable set of points over a finite field enables their use beyond traditional cryptographic applications. Building on these properties, we introduce \textit{ECM}, which utilizes elliptic curve point mappings to construct a modulation framework that ensures both symbol obfuscation and structured constellation formation. Specifically, in ECM, binary data is mapped onto distinct modulation symbols by exploiting the structured randomness of elliptic curve points. To achieve this, binary bit sequences are first segmented into fixed-length groups corresponding to the desired number of modulation symbols $M$. Each bit group of length $\log_2 M$ is then mapped to a set $K = \{k_i\}_{i=1}^{M}$ of unique integer scalars, which are subsequently used in the scalar multiplication operation $P_i = k_i G$, where $G$ is a predefined base point on the elliptic curve $E$ of large prime order $n$. This process generates a corresponding set of elliptic curve points $P = \{P_i\}_{i=1}^{M}$, ensuring a one-to-one correspondence between binary data and modulation symbols. The result is a modulation constellation consisting of $M$ distinct elliptic curve points, where the underlying elliptic curve properties naturally introduce a structured yet non-trivial mapping of information onto the signal space.

To maintain consistent transmission power, the generated constellation must be centered and normalized. This begins by computing the centroid $C$ of the constellation points in $P$ as
\begin{equation}
C = \left(\frac{\sum_{i=1}^{M}X_i}{M},  \frac{\sum_{i=1}^{M}Y_i}{M}\right),  
\end{equation}
where $\left(X_i, Y_i\right)$ represents the coordinates of each point in $P$. The constellation is then translated such that its centroid coincides with the origin, forming the adjusted set $P' = P - C$. Once centered, the constellation is further normalized to meet transmission energy constraints, ensuring practical implementation. The final set of symbols, denoted as $P'' = \{P''_i\}_{i=1}^{M}$, is obtained by scaling the constellation to achieve unit average energy. Each point in $P''$ is represented as a complex number $x_i + j y_i$, where $j$ is the imaginary unit and $\left(x_i, y_i\right)$ are the coordinates of the normalized constellation. This complex representation ensures compatibility with existing communication systems, enabling efficient signal processing and seamless integration with modern transceiver architectures.

An additional critical aspect of ECM design is ensuring a minimum Euclidean distance $\dmin$ between modulation symbols, as this directly influences the SEP performance. This is achieved by carefully selecting the scalar multiples and their corresponding elliptic curve points to enforce a minimum separation constraint within the final constellation $P''$. Since the base point $G$ has a large prime order $n$, the set $\{G,2\cdot G, \dots, n\cdot G\}$ forms an extensive, well-distributed collection of points on the elliptic curve \cite{elliptic-curve-book}. This structured lattice allows for the selection of $M$-tuples that satisfy the $\dmin$ requirement, ensuring robustness against noise while maintaining encryption strength. To formalize this, we define an $M$-ECM$(d_{\mathrm{min}})$ constellation as one consisting of $M$ points, each separated by a minimum distance $d_{\mathrm{min}}$.

\textbf{Example:} A $4$-ECM$(d_{\mathrm{min}})$ is considered, corresponding to two-bit information sequences $\{00, 01, 10, 11\}$. These sequences are mapped to a set of four different scalars $ K = \{k_1, k_2, k_3, k_4\}$, producing the elliptic curve points $\{P_1, P_2, P_3, P_4\}$ via the operation $P_i = k_i \cdot G$. The centroid $C$ of these points is computed, and each point is translated and normalized to ensure unit average energy, forming the final constellation:
\begin{equation}
   P'' = \{P_{1}'', P_{2}'', P_{3}'', P_{4}''\}.  
\end{equation}
The chosen scalars $\{k_i\}_{i=1}^{4}$ must be carefully selected to maintain the required minimum Euclidean distance $d_{\mathrm{min}}$ between any two points in $P''$. While this example illustrates the approach for $M=4$, the methodology generalizes to larger $M$-tuples, offering flexibility in adapting to varying performance constraints and supporting higher-order modulation schemes.

\begin{remark}
Although this work focuses on the single-user case, the ECM framework naturally scales to multi-user deployments by assigning each user a distinct secret key for generating independent $M$-tuple mappings offline, with the resulting increase in preprocessing cost having no impact on the constant-time modulation and demodulation operations, thereby preserving low-latency and bandwidth in large-scale networks.
\end{remark}

\section{ECM-based Physical Layer Encryption}
In this section, we present an ECM-based PLE framework that utilizes a key-based predefined $M$-tuple for each symbol transmission while maintaining a minimum distance constraint. Specifically, by ensuring that the selected $M$-tuples satisfy a specified $d_{\mathrm{min}}$, the proposed approach preserves SEP performance equivalent to unencrypted QAM systems while introducing varying constellation characteristics across transmissions to enhance data obfuscation. However, while ECM effectively achieves secure symbol mapping, its direct application to PLE introduces a computational challenge, as maintaining high security requires a sufficiently large set of valid $M$-tuples. To address this limitation, we develop a novel PLE scheme namely ECM-DR that enhances security beyond conventional dynamic constellation rotation techniques while maintaining the SEP performance of unencrypted systems, providing a robust and practical PLE mechanism.

\subsection{Data Obfuscation through ECM}

Unlike conventional modulation schemes that rely on fixed constellation mappings, ECM encodes transmitted symbols using elliptic curve transformations, leveraging their structured yet unpredictable properties to enhance security. By dynamically shaping constellation points through elliptic curves, ECM ensures that symbol representations remain concealed from an adversary, even when channel conditions are ideal. However, for this transformation to be effective in securing data transmission, it must be synchronized between legitimate users while maintaining robustness against inference attacks. To achieve this, Alice and Bob establish a shared secret key over a mutual or secure channel \cite{cryptographic-primitives}, which serves as a seed for a synchronized RNG accessible to both users. This seed enables the generation of a common set of valid $M$-tuples, each satisfying a predefined minimum distance constraint, ensuring that symbol mappings remain distinct and unpredictable across transmissions.

The role of these $M$-tuples extends beyond simple constellation shaping, forming the core of ECM’s obfuscation mechanism. Within this framework, Alice and Bob retain the first $N$ valid $M$-tuples that meet the $d_{\mathrm{min}}$ criterion, establishing a structured sequence to shuffle data and enhance obfuscation. By leveraging the deterministic order of these tuples, ECM guarantees that data transformation remains synchronized between the two legitimate users while ensuring that each transmission exhibits varying constellation characteristics. This continuous variation prevents an adversary from detecting stable patterns, significantly strengthening resilience against passive eavesdropping and statistical attacks. However, while elliptic curve structures inherently offer a large number of scalar sets that satisfy the $d_{\mathrm{min}}$ constraint \cite{elliptic-curve-book}, leveraging these sets requires careful selection, as arbitrary scalar choices may result in inefficient spacing between constellation points or redundant symbol mappings, which could impact both security and SEP performance. To mitigate these risks, ECM systematically extracts a diverse subset of valid $M$-tuples, ensuring that each transmitted constellation remains unique while maintaining the desired minimum distance properties. This controlled variation disrupts statistical predictability, preventing adversaries from recognizing patterns and significantly enhancing robustness against inference attacks. 

To achieve this in a structured and computationally efficient manner, Alice and Bob must implement a robust selection mechanism that preserves communication reliability while optimizing security. To this end, they first generate a sufficiently large set of scalars, $K = \{k_i\}_{i=1}^{L}$, where $L$ is chosen to ensure a wide distribution of candidate tuples. These scalars are then mapped to their corresponding elliptic curve points and undergo preprocessing, including centroid subtraction for centering and normalization to maintain unit average energy, ensuring a standardized basis for tuple construction. In this direction, to efficiently construct valid $M$-tuples, we employ Algorithm \ref{alg:m-tuples}, which iteratively selects scalars from the candidate pool while enforcing the minimum Euclidean distance constraint. The algorithm initializes an $M$-tuple by selecting a starting scalar and iteratively adding subsequent scalars whose elliptic curve points satisfy the predefined $d_{\mathrm{min}}$ requirement relative to previously selected points. To enforce this constraint efficiently, we utilize 2D trees \cite{k-d-trees} to accelerate nearest-neighbor searches, reducing computational overhead. Once an $M$-tuple is formed, its points are refined to preserve unit-average energy while ensuring that the constellation structure remains dynamic across transmissions. This structured yet adaptive selection process ensures that ECM maintains strong obfuscation capabilities while allowing Bob to decode the transmitted data despite continuous $M$-tuple variation reliably.

Next the asymptotic complexity of Algorithm~\ref{alg:m-tuples} is calculated.
\begin{proposition}
\label{prop:alg-cost}
Assuming the 2D‑tree in line 6 of Algorithm~\ref{alg:m-tuples} is built with the deterministic median splits \cite{kd_tree_build}, the time complexity of Algorithm~\ref{alg:m-tuples} is given by $
\mathcal{O}\!\Bigl(L\kappa + L\log L + NM\log L + NM^{2} \Bigr)$, where $L$, $M$, $N$, and $\kappa=\lceil\log_{2} n\rceil$ are
defined earlier.
\end{proposition}
\begin{IEEEproof}
The proof is provided in Appendix~\ref{appendix:Alg1_O_complexity}
\end{IEEEproof}

\begin{remark}
The complexity stated in Proposition~\ref{prop:alg-cost} applies exclusively to the offline preprocessing stage, where the kd-tree is constructed and the bank of valid $M$-tuples is synthesized. This step is performed once during initialization or upon key refresh, and lies entirely outside the real-time transmission path. During operation, each symbol is modulated via a constant-time table lookup and, in ECM-DR, a single complex rotation, both independent of the tuple generation. As a result, the online complexity remains $\mathcal{O}(1)$ per symbol, matching the efficiency of conventional $M$-QAM and ensuring that throughput and latency are unaffected by the offline cost.
\end{remark}

Finally, while the structured selection of $M$-tuples enhances security by continuously altering the transmitted constellation, it remains crucial to assess the extent of symbol obfuscation introduced by ECM. In more detail, an eavesdropper attempting to recover the transmitted symbols $s$ can analyze the received symbols $r$ over an extended period, leveraging statistical patterns to infer the underlying modulation scheme. To counteract this, ECM must not only prevent stable symbol mappings but also ensure that the scatter of received symbols in the IQ plane remains sufficiently unpredictable. This motivates the need for a quantifiable measure that captures the level of obfuscation achieved by ECM across transmissions. Thus, to evaluate this, we adopt an information-theoretic approach based on entropy, which quantifies the unpredictability of the received constellation. In this direction, since directly computing continuous entropy is impractical for digital systems that inherently quantize signals, we employ quantized information entropy as defined in \cite{cryptographic-primitives}:
\begin{equation}
    \begin{aligned}
        H^{\Delta}\left(r_x,r_y\right)\coloneqq -\sum_{i=-\infty}^{\infty}\sum_{j=-\infty}^{\infty}\Gamma\left(i,j\right)\log_{2}{\Gamma\left(i,j\right)},
    \end{aligned}
\end{equation}
where 
\begin{equation}
\Gamma\left(i,j\right) = \int_{i\Delta}^{(i+1)\Delta}\int_{j\Delta}^{(j+1)\Delta} p\left(r_x,r_y\right)dr_x dr_y,
\end{equation}
with $(r_x,r_y)$ denoting the coordinates of the received symbols, $\Delta$ representing the quantization accuracy, and $p(r_x, r_y)$ being the joint probability density function of $r_x$ and $r_y$. Notably, when $p(r_x, r_y)$ follows a 2D uniform distribution, the quantized information entropy $H^{\Delta}$ reaches its maximum, signifying an optimal level of unpredictability and security \cite{Elements-of-Information-Theory}.

\begin{algorithm}[!t]
\caption{Generating Valid \( M \)-Tuples for \( M \)-ECM(\(d_{\min}\))}
\label{alg:m-tuples}
\begin{algorithmic}[1]
\REQUIRE\( p \), \( a, b \), \( G \), \( n \), \( M \), \( d_{\min} \), \( N \),  \( \text{seed} \), \(\text{max\_attempts}\)
\ENSURE Set of valid \( M \)-tuples \( \mathcal{T} \)
\STATE Initialize random number generator with \( \text{seed} \)
\STATE Generate candidate scalars \( \{k_i\}_{i=1}^{L} \), where \( L \) is sufficiently large and \( k_i < n \)
\STATE Map scalars to points: \( P_i = k_i G \) for \( i = 1, \dots, L \)
\STATE Center points: \( P_i' = P_i - \frac{1}{L}\sum_{i=1}^{L} P_i \)
\STATE Scale to unit energy: \( P_i'' = \frac{P_i'}{\sqrt{\frac{1}{L}\sum_{i=1}^{L} \|P_i'\|^2}} \)
\STATE $\text{2D}\leftarrow\text{Build2DTree}(\{P_i''\})$
\STATE Initialize \( \mathcal{T} \leftarrow \emptyset \), \( a \leftarrow 0 \)
\WHILE{ \( |\mathcal{T}| < N' \) \textbf{and} \( a < \text{max\_attempts} \) }
    \STATE \( a \leftarrow a + 1 \)
    \STATE Initialize \( T \leftarrow \emptyset \), \( A \leftarrow \{1, \dots, L\} \)
    \STATE Randomly select \( k_{\text{start}} \) from \( A \), add to \( T \), remove from \( A \)
    \FOR{ \( m = 2 \) \TO \( M \) }
        \STATE Retrieve \( P_{\text{current}}'' \) from last scalar in \( T \)
        \STATE
           $S \leftarrow \text{2D}.\text{RangeQuery}(P_{\text{curr}}'',\,r=d_{\min},\,\text{mask}=A)$
        \IF{ \( S \) is empty }
            \STATE \textbf{break}
        \ENDIF
        \STATE Randomly select \( k_{\text{new}} \) from \( S \), add to \( T \), remove from \( A \)
    \ENDFOR
    \IF{ \( |T| = M \) }
        \STATE Compute centroid \( C_T = \frac{1}{M}\sum_{k \in T} P_k \)
        \STATE Center tuple points: \( P_k' = P_k - C_T \) for \( k \in T \)
        \STATE Scale tuple: \( P_k'' \leftarrow \frac{P_k''}{\sqrt{\frac{1}{M}\sum_{k \in T} \|P_k''\|^2}} \) for \( k \in T \)
        \IF{ all pairwise \( \|P_i'' - P_j''\| \geq d_{\min} \) for \( i \neq j \)}
            \STATE Add \( T \) to \( \mathcal{T} \)
        \ENDIF
    \ENDIF
\ENDWHILE
\RETURN \( \mathcal{T} \)
\end{algorithmic}
\end{algorithm}

\subsection{Proposed PLE Scheme}
As previously discussed, ECM provides high obfuscation capabilities by employing predefined $M$-tuples in each transmission that satisfy a minimum Euclidean distance constraint $d_{\mathrm{min}}$ without compromising SEP performance. However, as the number of transmitted symbols $N$ increases, the computational cost of generating valid $M$-tuples grows significantly. Since each transmission ideally requires a unique $M$-tuple to maximize security, a large candidate pool $L$ must be maintained to ensure a sufficient number of valid $M$-tuples. More specifically, as $N$ increases and $L$ remains constant, the probability of finding new $M$-tuples that satisfy $d_{\mathrm{min}}$ diminishes due to the finite number of valid selections within the elliptic curve field. Since each selection reduces the remaining pool of available tuples, it becomes increasingly difficult to meet the $d_{\mathrm{min}}$ constraint, requiring more computational effort to identify new valid tuples. One approach to mitigate this challenge is to increase the candidate pool size $L$, allowing for more possible selections; however, this comes at the cost of higher algorithmic complexity, as the search space expands, making the offline precomputation phase increasingly demanding, particularly for large-scale transmissions or resource-constrained systems.
\begin{figure} 
    \centering
    \includegraphics[width=\linewidth]{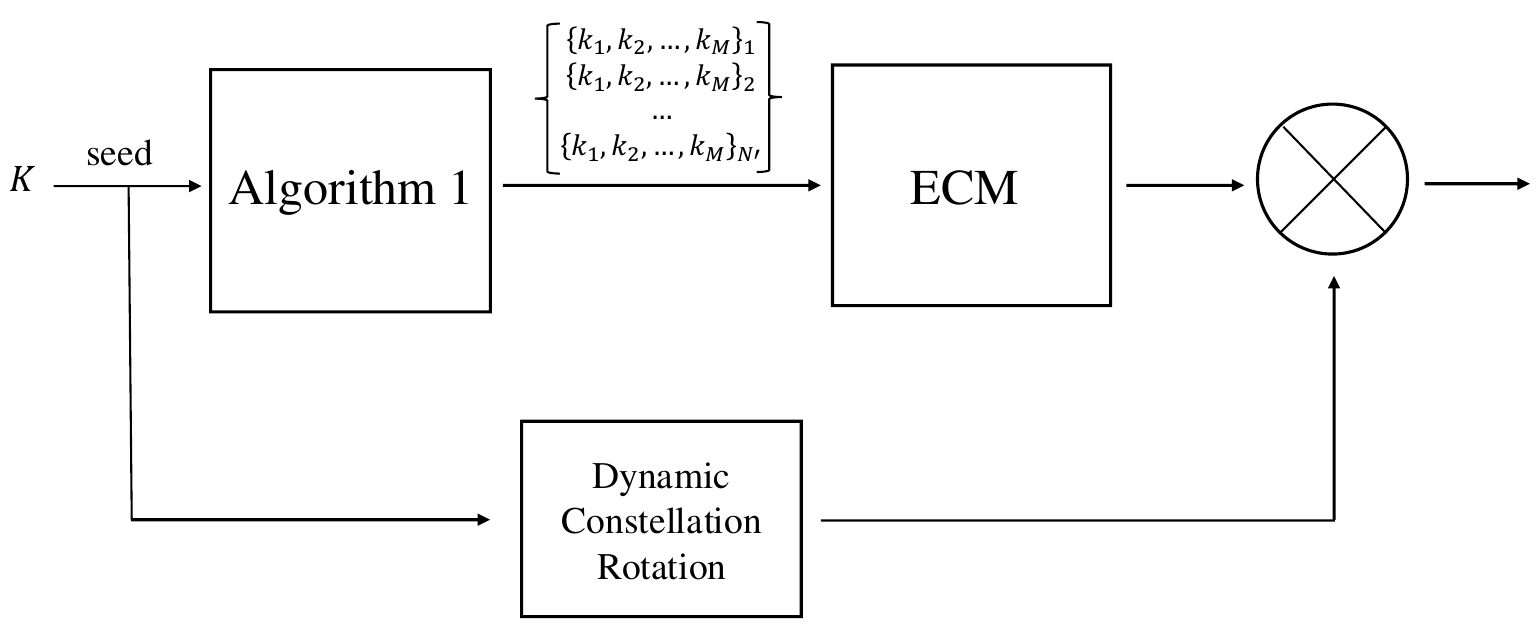} 
    \caption{Elliptic Curve Modulation with dynamic constellation rotation (ECM-DR) for physical layer encryption.}
    \label{fig:ECM-DR}
\end{figure}

To address this offline computational overhead, we propose ECM-DR, a practical PLE scheme that achieves near-maximal symbol obfuscation without requiring a unique $M$-tuple for each transmission. Instead of computing a new set of $M$-tuples for every symbol transmission, ECM-DR leverages a smaller subset of $N'\leq N$ valid $M$-tuples, significantly reducing offline computations by requiring only $N'$ rather than $N$ unique tuples. This strategic reduction in precomputed tuples alleviates the growing complexity associated with generating an excessively large set of valid $M$-tuples while preserving ECM’s encryption strength. Since reusing a smaller subset of $M$-tuples across transmissions introduces some degree of repetition, this could potentially lower information entropy compared to the ideal case where each transmission has a distinct $M$-tuple. To counteract this and maintain high levels of obfuscation, ECM-DR applies a lightweight, key-based dynamic constellation rotation in each transmission, as shown in Fig. \ref{fig:ECM-DR}. In more detail, the shared key $K$ serves as a seed for both Alice’s and Bob’s RNGs, ensuring synchronized encryption and decryption. This seed is used to initialize Algorithm 1, which generates $N'$ unique $M$-tuples, where $N'$ is kept small for computational feasibility, making offline precomputation manageable. Each $M$-tuple is then processed by the ECM module, producing $N'$ distinct constellation configurations. To further enhance security, a dynamic constellation rotation module generates pseudorandom phase rotations for each transmission, which is computationally lightweight and ensures that every transmission remains unique. The final modulated signal, now incorporating both ECM-based obfuscation and dynamic rotation, achieves a highly robust PLE scheme, significantly strengthening security against adversarial inference.

From a computational standpoint, ECM-DR reduces the total offline complexity of Algorithm \ref{alg:m-tuples}, which is given as
$\mathcal{O}\big(L\kappa + L\log L + N'M\log L + N'M^2\big)$, since only a smaller number of tuples, denoted as $N' \ll N$, is synthesized and stored. This reduced requirement also allows for a smaller candidate pool size $L$, which further lowers the preprocessing cost. During transmission, each symbol is generated through a constant-time table lookup, followed by a single complex rotation in the case of ECM-DR. This ensures that the modulation complexity remains independent of the tuple generation process. As a result, ECM-DR preserves the core security properties of ECM while significantly reducing offline computation, which makes it suitable for real-time and resource-constrained environments.

\section{Security Analysis}
The proposed ECM presents two primary ways in which an adversary, Eve, could attempt to compromise the confidentiality of the transmitted information either attacking the key generator or targeting the $M$-tuples used in the encryption process. Evaluating the complexity and feasibility of these brute-force attacks is essential to determine the scheme's robustness and identify potential bottlenecks in its security architecture.

\subsection{Attacks on Key Generation}
In this brute-force attack, Eve targets the key generator responsible for producing the shared random seed utilized by Alice and Bob. The security of this attack is directly tied to the strength of the key exchange protocol and the parameters of the seed stream cipher in use. Specifically, Eve may employ a brute-force approach, systematically searching through all possible $N_{s}$-bit keys to uncover the shared secret key. The computational complexity of this attack is exponential, with the search space comprising $2^{N_{s}}$ possible keys. For example, a $256$-bit seed results in $2^{256}$ potential keys, rendering such an exhaustive search computationally infeasible with current technology. This exponential growth in complexity ensures that provided $N_{s}$ is sufficiently large (e.g., 128 bits or more), brute-force attacks against the key generator remain practically unattainable. Consequently, the key generator's resistance to exhaustive search attacks serves as a formidable barrier, safeguarding the shared seed from unauthorized discovery and, by extension, maintaining the synchronization between Alice and Bob's 
$M$-tuple generation processes.

\subsection{Attacks on $M$-tuples}
In this brute-force attack, Eve seeks to bypass the key generator's security without directly attacking the seed, she would need to target the $M$-tuples themselves. However, this approach presents a significantly more complex and resource-intensive challenge compared to attacking the key generator. Without knowledge of the shared key, Eve must exhaustively search through all possible $M$-tuples to identify those that satisfy the minimum Euclidean distance $\dmin$ constraint. The computational complexity of this attack grows exponentially, as the number of possible $M$-tuples grows rapidly with the number of candidate scalars $L$. To quantitatively assess this, we present the following theorem, which estimates the expected number of valid $M$-tuples under a uniform two-dimensional distribution of points.

\begin{proposition}
\label{prop:prop2}
Given $L$ uniformly distributed points over a planar area $A$, the expected number $\mathbb{E}[T]$ of valid $M$-tuples that satisfy a minimum Euclidean distance $\dmin$ between all pairs of points within each tuple is approximately
\begin{equation} \label{eq:P_v}
    \begin{aligned}
        \mathbb{E}[T] \approx \frac{L^M}{M!} \exp\left(-\frac{\pi d_{\min}^2}{A}  \binom{M}{2}\right).
    \end{aligned}
\end{equation}

\end{proposition}

\begin{IEEEproof}
The proof is provided in Appendix~\ref{appendix:B}.
\end{IEEEproof}

 \begin{remark}
  Although Algorithm 1 is randomized, it converges reliably when the candidate pool size $L$ is chosen as a sufficient multiple of the desired number of tuples $N$, since the expected number of valid $M$-tuples grows exponentially with $L$, and convergence is enforced in practice through a maximum-attempts safeguard during tuple generation.
 \end{remark}

This approximation reveals an interplay between the parameters that influence the expected number of valid $M$-tuples. While the number of possible $M$-tuples scales combinatorially with $L$, the size of the candidate pool and the strict distance constraints imposed by $\dmin$ introduce an exponential decay factor that significantly reduces the number of valid tuples. The term $\frac{\pi\dmin^2}{A}$ in the exponential, where $A$ represents the planar search area, emphasizes that increasing $A$ mitigates the effect of the distance constraints and increases the likelihood of finding valid tuples. However, as the constellation order $M$ increases, the factorial term $M!$ introduced by $\binom{M}{2}$ dominates, causing the expected number of valid $M$-tuples to decrease exponentially. At the same time, the $\dmin$ constraint typically becomes less stringent for higher order constellations (e.g., $\dmin = 0.63$ for $16$-QAM and $\dmin = 0.3$ for $64$-QAM), partially offsetting the exponential decay by allowing more tuples to satisfy the relaxed constraint. This delicate balance ensures that even with higher order constellations, the expected number of valid $M$-tuples can remain substantial, maintaining the robustness of the scheme.

To illustrate the scale of the search space, consider a scenario with $M=16$, $\dmin=0.63$, $L=100000$, and $A=4$. Applying the approximation for the expected number of valid $M$-tuples yields a value on the order of $10^{50}$. This astronomically large value of possible valid $M$-tuples implies the vast combinatorial possibilities for constructing $M$-tuples that satisfy the given constraints, even under relatively stringent conditions. Such a result provides an intuitive grasp of the immense search space that Eve would have to navigate if she attempted an exhaustive search of the valid tuples. As a result, despite the exponential reduction introduced by $\dmin$ and $M$, the expected number of valid $M$-tuples remains astronomically large for practical values of $L$, $A$, and $M$. This abundance of valid tuples serves as a significant barrier for Eve, making it computationally infeasible to identify the specific tuples used by Alice and Bob through exhaustive search methods. The combination of a large candidate pool $L$ and a sufficiently large search area $A$ ensures that the tuples are numerous, and thus even without access to the shared seed, the overwhelming number of possibilities makes it virtually impossible for Eve to identify the $M$-tuples by brute-force attack, further enhancing the security robustness of the proposed scheme.

Within this context, it is also important to evaluate ECM’s resilience under the emergence of quantum adversaries. Unlike public key cryptographic protocols, ECM does not rely solely on the ECDLP for its symbol-level encryption. Instead, security arises from the exponentially large space of valid elliptic curve mappings with minimum distance constraints, which remain computationally infeasible to brute-force, even with quantum capabilities. Therefore, the primary quantum vulnerability lies in the key distribution mechanism used. However, since ECM is agnostic to the key exchange protocol, quantum-resistant approaches such as QKD or post-quantum key encapsulation mechanisms can be integrated to ensure end-to-end quantum-safe communication. This separation between key distribution and modulation further highlights ECM’s adaptability and robustness in post-quantum environments.

\section{Simulation Results}
This section evaluates the encryption effectiveness and reliability of ECM and its practical PLE implementation, ECM-DR, in terms of symbol obfuscation, information entropy, and SEP. We analyze how ECM enhances security by introducing symbol-level unpredictability while maintaining error performance equivalent to unencrypted QAM systems. Furthermore, we demonstrate how ECM-DR mitigates computational complexity while preserving strong encryption performance, making it a practical and robust PLE solution.
Finally, Table~\ref{tab:sim_params} summarizes the key parameters used in all simulation experiments presented in this section.

\begin{table}[h!]
	\centering
	\caption{Key Simulation Parameters}
	\label{tab:sim_params}
	\begin{tabular}{|l|l|}
		\hline
		\textbf{Parameter} & \textbf{Value} \\ \hline
		Modulation Schemes & ECM, ECM-DR, QAM-DR, QAM-SOTP \\ \hline
		Modulation Order ($M$) & 4, 16 \\ \hline
		Quantization Length ($q$) & 6–9 bits \\ \hline
		Subset Sizes ($N'$) & 50, 300 \\ \hline
	\end{tabular}
\end{table}

In Figs. \ref{fig:1a} and \ref{fig:1b}, the significant symbol obfuscation capabilities of ECM is illustrated, making constellation decoding highly challenging for an adversary. Unlike QPSK-DR, which applies rotational transformations, ECM fundamentally reshapes the constellation by leveraging elliptic curve point mappings. The results confirm that $4$-ECM($\sqrt{2}$) achieves superior obfuscation compared to QPSK-DR while maintaining identical SEP performance, as its minimum Euclidean distance $d_{\mathrm{min}} = \sqrt{2}$ matches that of conventional QPSK when normalized to unit average energy, preserving noise resilience. Beyond strong symbol obfuscation, ECM provides tunable encryption strength by adjusting $d_{\mathrm{min}}$, allowing to further enhance the security. More specifically, Figs. \ref{fig:1c} and \ref{fig:1d} illustrate how reducing $d_{\mathrm{min}}$ increases constellation density, enhancing unpredictability while still maintaining structured mappings, ensuring that legitimate receivers can reliably decode the signal while making interception infeasible for an eavesdropper.

\begin{figure}[h!]
\centering
\begin{subfigure}{0.45\columnwidth}
\centering
\begin{tikzpicture}[thick, scale=0.5]
\begin{axis}[
   xlabel={},
   ylabel={},
   xmajorticks=false,
   ymajorticks=false
]
\addplot+[
   only marks,
   mark options={black},
   mark=*, 
   mark size=3pt, 
] table {figures/Scatter_16_QAM_with_first1000_elements.txt};
\end{axis}
\end{tikzpicture}
\caption{16-QAM-DR} \label{fig:2a}
\end{subfigure}%
\hfill
\begin{subfigure}{0.45\columnwidth}
\centering
\begin{tikzpicture}[thick, scale=0.5]
\begin{axis}[
   xlabel={},
   ylabel={},
   xmajorticks=false,
   ymajorticks=false
]
\addplot+[
   only marks,
   mark options={black},
   mark=*, 
   mark size=3pt, 
] table {figures/Scatter_EC_wo_first1000_elements.txt};
\end{axis}
\end{tikzpicture}
\caption{16-ECM$\left(0.63\right)$} \label{fig:2b}
\end{subfigure}
\caption{Scatter diagram of noiseless 
$16$-QAM with dynamic constellation rotation and $16$-ECM$(0.63)$.}
\label{fig:2}
\end{figure}
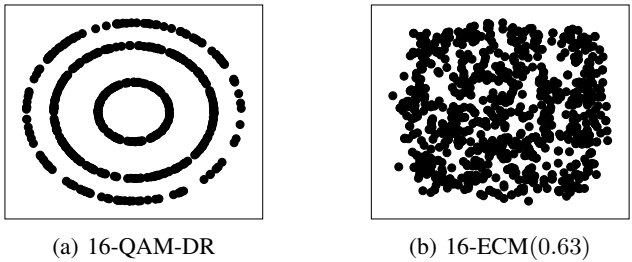

Fig.\ref{fig:2a} presents the scatter diagram of the $16$-QAM-DR scheme, where dynamic constellation rotation applies pseudorandom phase shifts to each symbol. While this operation effectively distorts angular information, it leaves the amplitude structure intact. As a result, the modulated symbols remain confined to concentric rings, revealing the original power levels of the underlying QAM constellation. This visible structure allows an eavesdropper to infer partial information, weakening the overall secrecy of the scheme. In contrast, Fig.\ref{fig:2b} shows the output of the $16$-ECM($0.63$) scheme, which constructs the constellation through entropy-maximizing mappings over elliptic curve points. Although it maintains the same minimum Euclidean distance as $16$-QAM to preserve SEP, the resulting distribution appears dense and structureless, effectively concealing both amplitude and phase. This demonstrates that ECM not only preserves communication performance but also offers substantially stronger protection against unauthorized signal analysis.

 Fig.~\ref{fig::bar_chart1} illustrates the information entropy achieved by $4$-ECM($\dmin$) and QPSK-DR under different minimum distance values and quantization resolutions. As shown, $4$-ECM($\sqrt{2}$) consistently outperforms QPSK-DR, achieving over 3.5 additional bits of entropy at each resolution. This advantage becomes more pronounced as quantization precision increases, since ECM utilizes the IQ plane more efficiently, leading to greater symbol dispersion and randomness. Moreover, reducing $d_{\mathrm{min}}$ further enhances entropy by allowing denser symbol packing, which drives the distribution toward uniformity. As a result, ECM approaches the theoretical entropy ceiling of $2$ times the quantization length, highlighting its capability to maximize obfuscation without sacrificing communication performance.
 
\begin{figure}[h!]
\centering
\begin{subfigure}{0.45\columnwidth}
\centering
\begin{tikzpicture}[thick, scale=0.5]
\begin{axis}[
   xlabel={},
   ylabel={},
   xmajorticks=false,
   ymajorticks=false
]
\addplot+[
   only marks,
   mark options={black},
   mark=*, 
   mark size=3pt, 
] table {figures/Received_Symbols_QPSK_with_rotation_SNR_50dB_dmin_1.txt};
\end{axis}
\end{tikzpicture}
\caption{QPSK-DR} \label{fig:1a}
\end{subfigure}%
\hfill
\begin{subfigure}{0.45\columnwidth}
\centering
\begin{tikzpicture}[thick, scale=0.5]
\begin{axis}[
   xlabel={},
   ylabel={},
   xmajorticks=false,
   ymajorticks=false
]
\addplot+[
   only marks,
   mark options={black},
   mark=*, 
   mark size=3pt, 
] table {figures/Received_Symbols_EC_without_rotation_SNR_50dB_dmin_1.38.txt};
\end{axis}
\end{tikzpicture}
\caption{4-ECM$\left(\sqrt{2}\right)$} \label{fig:1b}
\end{subfigure}%
\hfill
\begin{subfigure}{0.45\columnwidth}
\centering
\begin{tikzpicture}[thick, scale=0.5]
\begin{axis}[
   xlabel={},
   ylabel={},
   xmajorticks=false,
   ymajorticks=false
]
\addplot+[
   only marks,
   mark options={black},
   mark=*, 
   mark size=3pt, 
] table {figures/Received_Symbols_EC_without_rotation_SNR_50dB_dmin_1.2.txt};
\end{axis}
\end{tikzpicture}
\caption{4-ECM$\left(1.2\right)$} \label{fig:1c}
\end{subfigure}%
\hfill
\begin{subfigure}{0.45\columnwidth}
\centering
\begin{tikzpicture}[thick, scale=0.5]
\begin{axis}[
   xlabel={},
   ylabel={},
   xmajorticks=false,
   ymajorticks=false
]
\addplot+[
   only marks,
   mark options={black},
   mark=*, 
   mark size=3pt, 
] table {figures/Received_Symbols_EC_without_rotation_SNR_50dB_dmin_1.txt};
\end{axis}
\end{tikzpicture}
\caption{4-ECM$\left(1.0\right)$} \label{fig:1d}
\end{subfigure}
\caption{Scatter diagram of noiseless dynamically rotated QPSK and 4-ECM$\left(d_{\mathrm{min}}\right)$.}
\label{fig:1}
\end{figure}
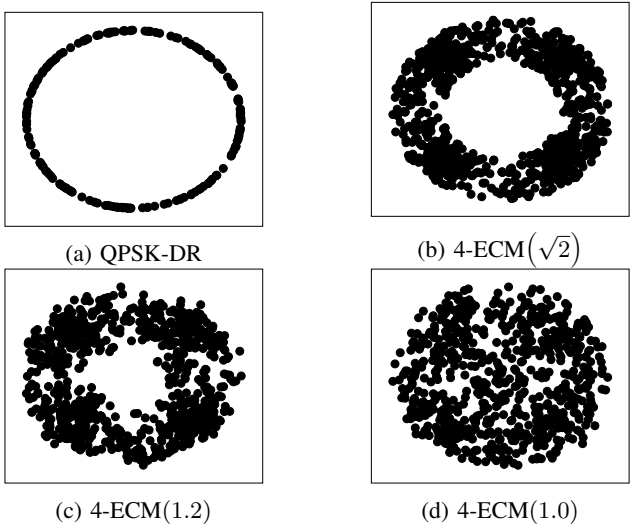

While ECM achieves strong encryption, its practical implementation requires managing precomputation complexity, which depends on the number of unique $M$-tuples. Figs. \ref{fig:3a} and \ref{fig:3b} illustrate that using a limited number of $M$-tuples ($N' = 50$) modestly reduces symbol obfuscation, as fewer unique mappings are available, leading to tuple reuse. However, this drastically reduces offline computational complexity, making ECM more viable for real-world deployment. To compensate for $M$-tuple reuse while preserving security, ECM-DR integrates key-based dynamic constellation rotation, a simple yet highly effective method that can be applied on top of ECM at each transmission. By leveraging ECM’s structured obfuscation capabilities, ECM-DR extends ECM into a practical and scalable PLE solution that balances security, computational efficiency, and error performance. Figures \ref{fig:3c} and \ref{fig:3d} demonstrate that ECM-DR successfully restores ECM’s obfuscation capability, effectively disrupting recognizable patterns introduced by tuple reuse while maintaining computational efficiency. This ensures that ECM-DR remains a scalable and practical encryption strategy, achieving near-optimal symbol obfuscation with significantly lower computational overhead, making it a robust PLE approach for real-world applications.

\begin{figure}[ht!]
    \centering
\begin{tikzpicture}
  \centering
  \begin{axis}[
        ybar,
        width=\columnwidth,
        bar width=0.15cm,
        ymajorgrids, tick align=inside,
        ymin=0, ymax=18,
        ytick={0,2,...,18},
        axis x line*=bottom,
        axis y line*=left,
        y axis line style={opacity=0},
        tickwidth=0pt,
        enlarge x limits=true,
        legend style={
            at={(0,1.18)},
            anchor=north west,
            legend columns=1,
            font = \tiny
        },
        ylabel={Information entropy (bit)},
        symbolic x coords={$6$ bit,$7$ bit,$8$ bit,$9$ bit},
        xtick=data,
       xlabel= $\text{Quantization length}$
    ]
     
    \addplot [draw=none, fill=yellow] coordinates {
    	($6$ bit, 8.34)
    	($7$ bit, 9.4)
    	($8$ bit, 10.35)
    	($9$ bit, 11.5)};
    \addplot [draw=none, fill=black] coordinates {
    	($6$ bit, 9.67)
    	($7$ bit, 11.64)
    	($8$ bit, 13.62)
    	($9$ bit, 15.4)};
    \addplot [draw=none,fill=red] coordinates {
    	($6$ bit, 9.35)
    	($7$ bit, 9.55)
    	($8$ bit, 9.61)
    	($9$ bit, 9.63)};
    \addplot [draw=none, fill=green] coordinates {
    	($6$ bit, 10.6)
    	($7$ bit, 11.68)
    	($8$ bit, 12)
    	($9$ bit, 12.18)};
    
    \addplot [draw=none, fill=brown] coordinates {
    	($6$ bit, 11.4)
    	($7$ bit, 13.3)
    	($8$ bit, 15)
    	($9$ bit, 16.7)};
    \addplot [draw=none, fill=blue] coordinates {
    	($6$ bit, 11.6)
    	($7$ bit, 13.4)
    	($8$ bit, 15.2)
    	($9$ bit, 17)};
    \legend{$16$-QAM-DR, $16$-QAM-SOTP, $16$-ECM$(0.63)$ $N'=50$, $16$-ECM$(0.63)$ $N'=300$, $16$-ECM$(0.63)$-DR $N'=50$, $16$-ECM$(0.63)$-DR $N'=300$}
  \end{axis}
  \end{tikzpicture}
  \captionsetup{justification=raggedright,singlelinecheck=false}
    \caption{Information entropy of $16$-ECM$(0.63)$-DR, $16$-QAM-DR and $16$-QAM-SOTP for various quantization lengths}
    \label{fig::bar_chart_hybrid}
\end{figure}
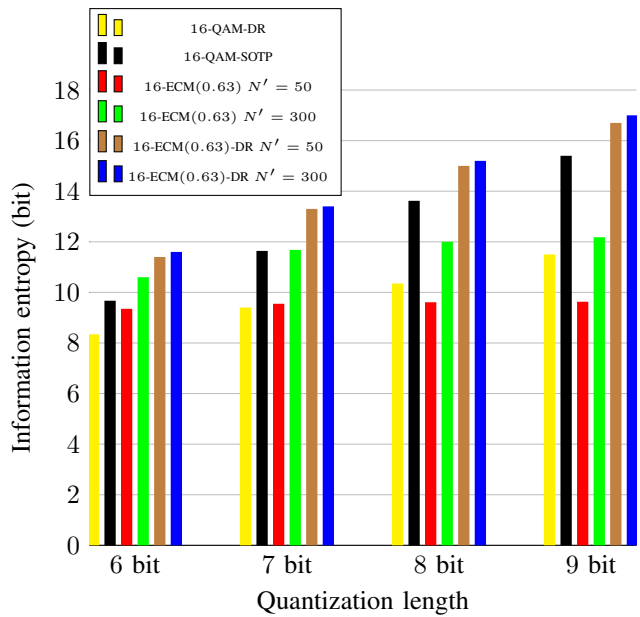

Figure~\ref{fig::bar_chart_hybrid} presents the information entropy achieved by the proposed $16$-ECM($0.63$)-DR scheme compared to baseline methods, including $16$-QAM-DR and $16$-QAM-SOTP, across quantization resolutions ranging from 6 to 9 bits. As shown, ECM-DR consistently outperforms QAM-DR and SOTP for both subset sizes $N'=50$ and $N'=300$, achieving entropy gains that increase with quantization precision. For $N'=50$, ECM-DR yields over 4 bits of additional entropy relative to QAM-DR and more than 2 bits over SOTP at 9-bit resolution. When $N'$ increases to $300$, these gains exceed 5 bits, indicating that ECM-DR effectively retains the high entropy characteristics of full ECM while reducing the computational cost of tuple synthesis. The figure also illustrates the entropy limits defined by the number of available constellation points, which are $800$ and $4800$ for $N'=50$ and $N'=300$, respectively. As quantization precision grows, ECM-DR approaches these bounds, demonstrating its ability to construct near-uniform symbol distributions. In contrast, SOTP exhibits saturation behavior due to its Gaussian-like symbol scatter, which restricts entropy maximization regardless of quantization resolution. Thus, by combining deterministic mappings with dynamic rotation, ECM-DR maintains strong obfuscation while reducing offline complexity, offering a practical balance between security and efficiency for real-world deployment.

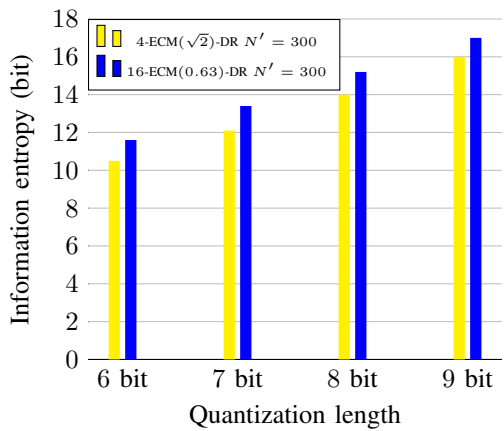
\begin{figure}[ht!]
	\centering
	\begin{tikzpicture}
		\centering
		\begin{axis}[
			ybar,
			width=0.8\columnwidth,
			bar width=0.15cm,
			ymajorgrids, tick align=inside,
			ymin=0, ymax=18,
			ytick={0,2,...,18},
			axis x line*=bottom,
			axis y line*=left,
			y axis line style={opacity=0},
			tickwidth=0pt,
			enlarge x limits=true,
			legend style={
				at={(0,1)},
				anchor=north west,
				legend columns=1,
				font = \tiny
			},
			ylabel={Information entropy (bit)},
			symbolic x coords={$6$ bit,$7$ bit,$8$ bit,$9$ bit},
			xtick=data,
			xlabel= $\text{Quantization length}$
			]
			
			\addplot [draw=none, fill=yellow] coordinates {
				($6$ bit, 10.5)
				($7$ bit, 12.1)
				($8$ bit, 14)
				($9$ bit, 16)};
			\addplot [draw=none,fill=blue] coordinates {
				($6$ bit, 11.6)
				($7$ bit, 13.4)
				($8$ bit, 15.2)
				($9$ bit, 17)};
			\legend{$4$-ECM$(\sqrt{2})$-DR $N'=300$, $16$-ECM$(0.63)$-DR $N'= 300$}
		\end{axis}
	\end{tikzpicture}
	\captionsetup{justification=raggedright,singlelinecheck=false}
	\caption{Information entropy of ECM-DR across various $M$}
	\label{fig::bar_chart_hybrid_M}	
\end{figure}

To further investigate the influence of system parameters on encryption strength, Fig.~\ref{fig::bar_chart_hybrid_M} presents the information entropy achieved by ECM-DR for different tuple sizes $M$, under a fixed scalar pool size of $N' = 300$. Specifically, the figure compares $4$-ECM($\sqrt{2}$)-DR and $16$-ECM($0.63$)-DR across quantization lengths ranging from 6 to 9 bits. As shown, increasing $M$ consistently results in higher entropy at all quantization resolutions. This trend stems from the fact that larger $M$ values allow more bits to be jointly encoded into a single symbol, enabling a richer spread of points across the complex plane. The resulting constellations exhibit greater density and uniformity, reducing symbol predictability and enhancing obfuscation. Importantly, while higher $M$ values increase the complexity of the offline tuple generation process, due to stricter minimum distance constraints, the real-time modulation procedure remains unaffected, relying only on precomputed lookups and lightweight complex rotations. This observation confirms that ECM-DR provides a tunable design space: by adjusting $M$, system designers can balance preprocessing effort against encryption strength, tailoring the scheme to suit diverse performance and security requirements.

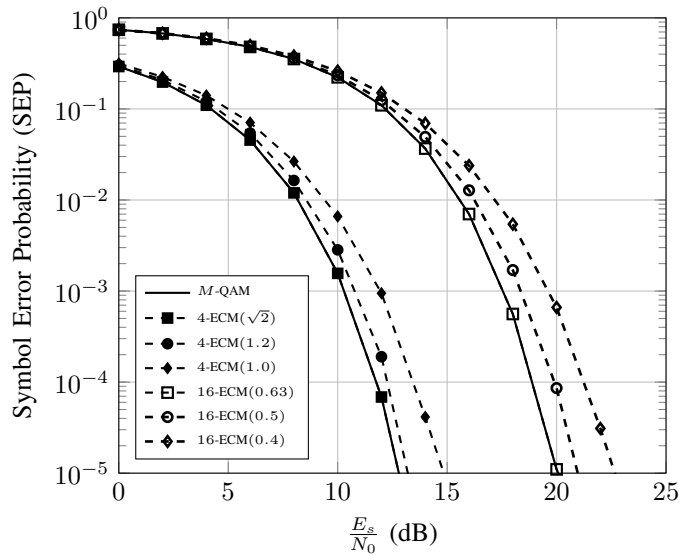
\begin{figure}
	\centering
	\begin{tikzpicture}
	\begin{semilogyaxis}[
    width=1\linewidth,
	xlabel = $\frac{E_{s}}{N_{0}}$ (dB),
	ylabel = Symbol Error Probability (SEP),
	xmin = 0,
	xmax = 25,
	ymin = 0.00001,
	ymax = 1,
    xtick = {0, 5,...,40},
	grid = major,
	legend cell align = {left},
    legend pos = south west,
    legend style={font=\tiny}
	]

	\addplot[
	no marks,
        color=black,
        line width = 0.8pt,
	style = solid,
	]
	table {figures/simulation_results/SER_vs_SNR_QPSK.txt};
        \addlegendentry{$M$-QAM};
        
        \addplot[
	style = dashed,       
        line width = 0.8pt,
        mark=square*,  
        mark options={solid},
        mark size = 1.8,
        color=black    
	]
	table {figures/simulation_results/SER_vs_SNR_EC_dmin_1.38.txt};
        \addlegendentry{$4$-ECM$(\sqrt{2})$};
        
        \addplot[
	style = dashed,       
        line width = 0.8pt,
        mark=*,  
        mark options={solid},
        mark size = 1.8,
        color=black    
	]
	table {figures/simulation_results/SER_vs_SNR_EC_dmin_1.2.txt};
        \addlegendentry{$4$-ECM$(1.2)$};

        \addplot[
	style = dashed,       
        line width = 0.8pt,
        mark=diamond*,  
        mark size = 1.8,
        mark options={solid},
        color=black    
	]
	table {figures/simulation_results/SER_vs_SNR_EC_dmin_1.txt};
        \addlegendentry{$4$-ECM$(1.0)$};


        \addplot[
	style = dashed,       
        line width = 0.8pt,
        mark=square,  
        mark options={solid},
        mark size = 2,
        color=black    
	]
	table {figures/simulation_results/SER_vs_SNR_16QAM.txt};
        \addlegendentry{$16$-ECM$(0.63)$};

        \addplot[
	style = dashed,       
        line width = 1pt,
        mark=o,  
        mark options={solid},
        mark size = 1.8,
        color=black    
	]
	table {figures/simulation_results/SER_vs_SNR_EC_dmin_0.5.txt};
        \addlegendentry{$16$-ECM$(0.5)$};

        \addplot[
	style = dashed,       
        line width = 1pt,
        mark=diamond,  
        mark options={solid},
        mark size = 2,
        color=black    
	]
	table {figures/simulation_results/SER_vs_SNR_EC_dmin_0.4.txt};
        \addlegendentry{$16$-ECM$(0.4)$};

        \addplot[
	no marks,
        color=black,
        line width = 0.8pt,
	style = solid,
	]
	table {figures/simulation_results/SER_vs_SNR_16QAM.txt};

	\end{semilogyaxis}
	\end{tikzpicture}
	\caption{SER of the legitimate user versus SNR}
	\label{fig::ser}
\end{figure}

Finally, Fig. \ref{fig::ser} compares the SEP performance of ECM-based constellations across various $\dmin$ values to unencrypted QAM. The results confirm that $4$-ECM($\sqrt{2}$) and $16$-ECM($0.63$) maintain identical SEP to QPSK and $16$-QAM, respectively, as their $d_{\mathrm{min}}$ values match those of unencrypted QAM when normalized to unit average energy, ensuring equivalent noise resilience. Furthermore, ECM allows tunability by adjusting $\dmin$, increasing information entropy at the cost of a higher SNR requirement. The results indicate that $4$-ECM($1.2$) and $4$-ECM($1.0$) require approximately 0.8 dB and 2.5 dB higher SNR, respectively, than QPSK, while $16$-ECM(0.5) and $16$-ECM($0.4$) require approximately 1.5 dB and 3 dB higher SNR, respectively, than $16$-QAM. Since ECM-DR inherits ECM’s symbol obfuscation properties, it maintains the same SEP while reducing computational complexity. By integrating key-based dynamic constellation rotation, ECM-DR preserves strong encryption performance without compromising reliability.

\begin{figure}[ht!]
    \centering
\begin{tikzpicture}
  \centering
  \begin{axis}[
        ybar,
        width=0.97\columnwidth,
        bar width=0.15cm,
        ymajorgrids, tick align=inside,
        ymin=0, ymax=18,
        ytick={0,2,...,18},
        axis x line*=bottom,
        axis y line*=left,
        y axis line style={opacity=0},
        tickwidth=0pt,
        enlarge x limits=true,
        legend style={
            at={(0,1.12)},
            anchor=north west,
            legend columns=1,
            font = \tiny
        },
        ylabel={Information entropy (bit)},
        symbolic x coords={$6$ bit,$7$ bit,$8$ bit,$9$ bit},
        xtick=data,
       xlabel= $\text{Quantization length}$
    ]
      
    \addplot [draw=none, fill=yellow] coordinates {
      ($6$ bit, 7.35)
      ($7$ bit, 8.36)
      ($8$ bit, 9.57)
      ($9$ bit, 11)};
   \addplot [draw=none,fill=red] coordinates {
      ($6$ bit, 10.5)
      ($7$ bit, 12.1)
      ($8$ bit, 14)
      ($9$ bit, 16)};
   \addplot [draw=none, fill=green] coordinates {
      ($6$ bit, 10.9)
      ($7$ bit, 12.61)
      ($8$ bit, 14.4)
      ($9$ bit, 16.3)};
   \addplot [draw=none, fill=brown] coordinates {
      ($6$ bit, 11.22)
      ($7$ bit, 12.82)
      ($8$ bit, 14.78)
      ($9$ bit, 16.78)};
    \legend{QPSK-DR, $4$-ECM$(\sqrt{2})$, $4$-ECM$(1.2)$, $4$-ECM$(1.0)$}
  \end{axis}
  \end{tikzpicture}
  \captionsetup{justification=raggedright,singlelinecheck=false}
    \caption{Information entropy of $4$-ECM$(\dmin)$ and QPSK-DR for various quantization lengths}
    \label{fig::bar_chart1}
\end{figure}
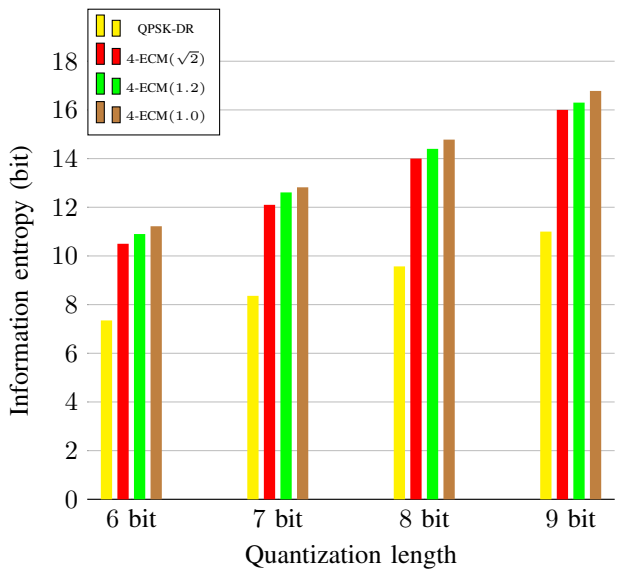

\begin{figure}[h!]
\centering
\begin{subfigure}{0.45\columnwidth}
\centering
\begin{tikzpicture}[thick, scale=0.5]
\begin{axis}[
   xlabel={},
   ylabel={},
   xmajorticks=false,
   ymajorticks=false
]
\addplot+[
   only marks,
   mark options={black},
   mark=*, 
   mark size=3pt, 
] table {figures/hybrid/Received_Symbols_EC_without_rotation_SNR_50dB_dmin_1.38.txt};
\end{axis}
\end{tikzpicture}
\caption{$4$-ECM$(\sqrt{2})$ without dynamic rotation} \label{fig:3a}
\end{subfigure}%
\hfill
\begin{subfigure}{0.45\columnwidth}
\centering
\begin{tikzpicture}[thick, scale=0.5]
\begin{axis}[
   xlabel={},
   ylabel={},
   xmajorticks=false,
   ymajorticks=false
]
\addplot+[
   only marks,
   mark options={black},
   mark=*, 
   mark size=3pt, 
] table {figures/hybrid/Scatter_EC_without_rotation_SNR_50dB_dmin_0.55.txt};
\end{axis}
\end{tikzpicture}
\caption{$16$-ECM$(0.63)$ without dynamic rotation} \label{fig:3b}
\end{subfigure}%
\hfill
\begin{subfigure}{0.45\columnwidth}
\centering
\begin{tikzpicture}[thick, scale=0.5]
\begin{axis}[
   xlabel={},
   ylabel={},
   xmajorticks=false,
   ymajorticks=false
]
\addplot+[
   only marks,
   mark options={black},
   mark=*, 
   mark size=3pt, 
] table {figures/hybrid/Received_Symbols_EC_with_rotation_SNR_50dB_dmin_1.38.txt};
\end{axis}
\end{tikzpicture}
\caption{$4$-ECM$(\sqrt{2})$ with dynamic rotation} \label{fig:3c}
\end{subfigure}%
\hfill
\begin{subfigure}{0.45\columnwidth}
\centering
\begin{tikzpicture}[thick, scale=0.5]
\begin{axis}[
   xlabel={},
   ylabel={},
   xmajorticks=false,
   ymajorticks=false
]
\addplot+[
   only marks,
   mark options={black},
   mark=*, 
   mark size=3pt, 
] table {figures/hybrid/Scatter_EC_with_rotation_SNR_50dB_dmin_0.55.txt};
\end{axis}
\end{tikzpicture}
\caption{$16$-ECM$(0.63)$ with dynamic rotation} \label{fig:3d}
\end{subfigure}%
\caption{Scatter diagram of noiseless $4$-ECM$\left(\sqrt{2}\right)$ and $16$-ECM$(0.63)$ with and without dynamic constellation rotation for $N'=20$ and $N=1000$.}
\label{fig:3}
\end{figure}
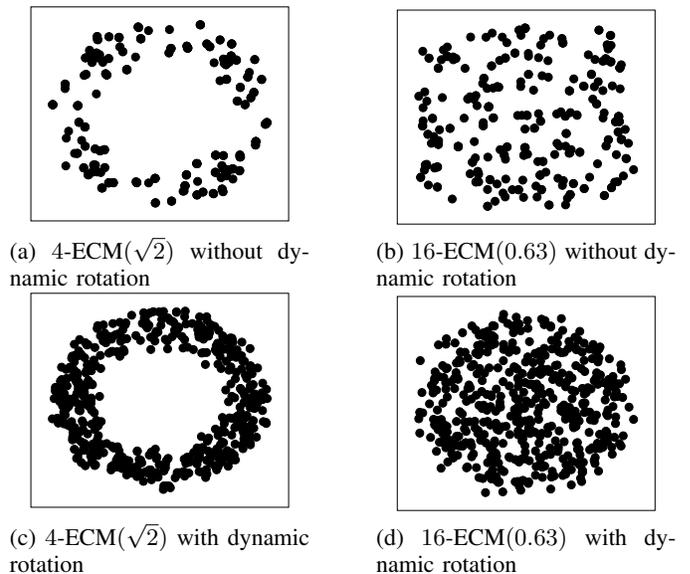

\color{black}

\vspace{-2mm}
\section{Conclusion}
This paper introduced Elliptic Curve Modulation (ECM), a novel symbol-level obfuscation framework that leverages the structured distribution of elliptic curve points to achieve near-optimal entropy without compromising SEP performance. By combining ECM with dynamic constellation rotation, we proposed ECM-DR, a practical physical-layer encryption scheme that enhances security while significantly reducing offline complexity. The design supports constant-time modulation per symbol, ensuring real-time efficiency and scalability across a wide range of system settings. Simulation results confirmed that ECM-DR consistently outperforms state-of-the-art PLE methods, including QAM-DR and SOTP, in terms of information entropy, while maintaining robustness and communication reliability. Future work will explore further optimization of the tuple generation process, extensions to multi-user and MIMO systems, and the integration of ECM’s structure into joint encryption and error correction designs. Additionally, refining bit-to-symbol mappings tailored to ECM constellations may further enhance BER performance without affecting symbol-level guarantees.

\section*{Acknowledgment}
The authors would like to thank Dimitrios Tyrovolas, Sotirios K. Michos, Sotiris A. Tegos, and Panagiotis D. Diamantoulakis for their valuable discussions and insights that contributed to the development of this work. Their feedback and suggestions helped refine key aspects of this study.

\appendices
\section{Proof of Proposition \ref{prop:alg-cost}} 
\label{appendix:Alg1_O_complexity}
The overall time complexity of Algorithm \ref{alg:m-tuples} is obtained by summing the costs of its three main phases: scalar preprocessing, kd-tree construction, and tuple generation. In the first phase, i.e., lines 2–5, the algorithm maps each of the $L$ candidate scalars to an elliptic curve point using scalar multiplication. This is performed using the standard double-and-add method, which requires $\mathcal{O}(\kappa)$ group operations per scalar, where $\kappa=\lceil\log_{2} n\rceil$ is the bit-length of the group order. The total cost is therefore $\mathcal{O}(L\kappa)$. The subsequent centering and energy normalisation steps involve linear-time passes over the point set, adding $\mathcal{O}(L)$, which is asymptotically dominated by the scalar multiplication term.  In the second phase, i.e., line 6, the algorithm builds a two-dimensional 2d-tree over the centred and normalised points using recursive median splits. This yields a balanced tree of height $\mathcal{O}(\log L)$ with worst-case construction time  $\mathcal{O}(L\log L)$ \cite{kd_tree_build}. The third phase, i.e., lines 11–29, synthesises $N$ valid  $M$-tuples. For each tuple, the algorithm selects a seed point, then iteratively performs $M-1$ radius queries to find points satisfying the minimum-distance constraint. Each such query traverses the 2d-tree in $\mathcal{O}(\log L)$ time, and each iteration involves at most $M$ pairwise distance checks. As a result, the cost per $M$-tuple is $\mathcal{O}(M\log L + M^2)$, and generating all $N$ tuples requires  $\mathcal{O}(NM\log L + NM^2)$ time. Combining the three phases completes the proof of Proposition \ref{prop:alg-cost}.

\section{Proof of Proposition \ref{prop:prop2}} 
\label{appendix:B}
Consider $L$ points uniformly distributed over a planar area 
$A$. The goal is to determine the expected number of $M$-tuples where every pair of points within each tuple is at least $\dmin$ apart. First, the probability $P_v$ that any two points are at least $\dmin$ apart is derived by considering the exclusion zone around each point. The exclusion area around a single point is $\pi\dmin^2$, representing the region within which no other point can lie to satisfy the distance constraint. Therefore, the probability that a randomly selected pair of points satisfies the distance constraint is expressed as
\begin{equation}
	\begin{aligned}
		P_v = 1-\frac{\pi\dmin^2}{A}.
	\end{aligned}
\end{equation}
An $M$-tuple consists of $\binom{M}{2}$ unique pairs of points thus the probability that all pairwise distances within the tuple exceed $\dmin$ is $P_{v}^{\binom{M}{2}}$. The total number of possible $M$-tuples from $L$ candidates is given by the binomial coefficient $\binom{L}{M}$, thereby the expected number of valid $M$-tuples is expressed as
\begin{equation} \label{eq:T}
	\begin{aligned}
		\mathbb{E}[T] = \binom{L}{M} \left( 1-\frac{\pi\dmin^2}{A} \right)^{\binom{M}{2}}.
	\end{aligned}
\end{equation}
Given that $L$ is sufficiently large and $M$, representing the constellation order, assumes moderate values, and since $\frac{\pi\dmin^2}{A}\ll 1$ due to the search planar area $A$ being sufficiently larger than $\dmin^2$, \eqref{eq:T} can be reformulated as follows
\begin{equation}
	\begin{aligned}
		\mathbb{E}[T] \approx \frac{L^M}{M!} \exp{\left(-\frac{\pi d_{\min}^2}{A}  \binom{M}{2}\right)},
	\end{aligned}
\end{equation}
which concludes the proof.

\bibliographystyle{IEEEtran}
\bibliography{Bibliography}

\begin{thebibliography}{10}
\providecommand{\url}[1]{#1}
\csname url@samestyle\endcsname
\providecommand{\newblock}{\relax}
\providecommand{\bibinfo}[2]{#2}
\providecommand{\BIBentrySTDinterwordspacing}{\spaceskip=0pt\relax}
\providecommand{\BIBentryALTinterwordstretchfactor}{4}
\providecommand{\BIBentryALTinterwordspacing}{\spaceskip=\fontdimen2\font plus
\BIBentryALTinterwordstretchfactor\fontdimen3\font minus
  \fontdimen4\font\relax}
\providecommand{\BIBforeignlanguage}[2]{{%
\expandafter\ifx\csname l@#1\endcsname\relax
\typeout{** WARNING: IEEEtran.bst: No hyphenation pattern has been}%
\typeout{** loaded for the language `#1'. Using the pattern for}%
\typeout{** the default language instead.}%
\else
\language=\csname l@#1\endcsname
\fi
#2}}
\providecommand{\BIBdecl}{\relax}
\BIBdecl

\bibitem{6G}
Z.~Zhang, Y.~Xiao, Z.~Ma, M.~Xiao, Z.~Ding, X.~Lei, G.~K. Karagiannidis, and
  P.~Fan, ``6g wireless networks: Vision, requirements, architecture, and key
  technologies,'' \emph{IEEE Veh. Technol. Mag.}, vol.~14, no.~3, pp. 28--41,
  2019.

\bibitem{SuperConstellations}
T.~K. Oikonomou, D.~Tyrovolas, S.~A. Tegos, P.~D. Diamantoulakis,
  P.~Sarigiannidis, and G.~K. Karagiannidis, ``On the design of super
  constellations,'' \emph{IEEE Open J. Commun. Soc.}, vol.~6, pp. 2741--2756,
  2025.

\bibitem{pls1}
Y.-S. Shiu, S.~Y. Chang, H.-C. Wu, S.~C.-H. Huang, and H.-H. Chen, ``Physical
  layer security in wireless networks: a tutorial,'' \emph{IEEE Wireless
  Commun.}, vol.~18, no.~2, pp. 66--74, 2011.

\bibitem{forensics1}
Z.~Sheng, H.~D. Tuan, A.~A. Nasir, H.~V. Poor, and E.~Dutkiewicz, ``Physical
  layer security aided wireless interference networks in the presence of strong
  eavesdropper channels,'' \emph{IEEE Trans. Inf. Forensics Security}, vol.~16,
  pp. 3228--3240, 2021.

\bibitem{forensics2}
R.~Chopra, C.~R. Murthy, and R.~Annavajjala, ``Physical layer security in
  wireless sensor networks using distributed co-phasing,'' \emph{IEEE Trans.
  Inf. Forensics Security}, vol.~14, no.~10, pp. 2662--2675, 2019.

\bibitem{pls2}
A.~Mukherjee, S.~A.~A. Fakoorian, J.~Huang, and A.~L. Swindlehurst,
  ``Principles of physical layer security in multiuser wireless networks: A
  survey,'' \emph{IEEE Commun. Surveys Tuts.}, vol.~16, no.~3, pp. 1550--1573,
  2014.

\bibitem{pls3}
Y.~Wu, A.~Khisti, C.~Xiao, G.~Caire, K.-K. Wong, and X.~Gao, ``A survey of
  physical layer security techniques for {5G} wireless networks and challenges
  ahead,'' \emph{IEEE J. Sel. Areas Commun.}, vol.~36, no.~4, pp. 679--695,
  2018.

\bibitem{thanos}
\BIBentryALTinterwordspacing
A.~P. Chrysologou, N.~D. Chatzidiamantis, A.-A.~A. Boulogeorgos, and Z.~Ding,
  ``On the reliability and security of ambient backscatter uplink {NOMA}
  networks,'' 2024. [Online]. Available: \url{https://arxiv.org/abs/2405.07057}
\BIBentrySTDinterwordspacing

\bibitem{apostolos}
A.~A. Tegos, Y.~Xiao, S.~A. Tegos, G.~K. Karagiannidis, and P.~D.
  Diamantoulakis, ``Trustworthy slotted {ALOHA},'' \emph{IEEE Wireless Commun.
  Lett.}, vol.~13, no.~12, pp. 3400--3403, 2024.

\bibitem{pls4}
W.~Trappe, ``The challenges facing physical layer security,'' \emph{IEEE
  Commun. Mag.}, vol.~53, no.~6, pp. 16--20, 2015.

\bibitem{covert1}
X.~Chen, J.~An, Z.~Xiong, C.~Xing, N.~Zhao, F.~R. Yu, and A.~Nallanathan,
  ``Covert communications: A comprehensive survey,'' \emph{IEEE Commun. Surv.
  Tutor.}, vol.~25, no.~2, pp. 1173--1198, 2023.

\bibitem{ple1}
R.~Ma, L.~Dai, Z.~Wang, and J.~Wang, ``Secure communication in {TDS}-{OFDM}
  system using constellation rotation and noise insertion,'' \emph{IEEE Trans.
  on Cons. Electron.}, vol.~56, no.~3, pp. 1328--1332, 2010.

\bibitem{ple2}
A.~Zuquete and J.~Barros, ``Physical-layer encryption with stream ciphers,'' in
  \emph{2008 IEEE International Symposium on Information Theory}, 2008, pp.
  106--110.

\bibitem{ple3}
C.~Zhang, J.~Yue, L.~Jiao, J.~Shi, and S.~Wang, ``A novel physical layer
  encryption algorithm for {LoRa},'' \emph{IEEE Communications Letters},
  vol.~25, no.~8, pp. 2512--2516, 2021.

\bibitem{ple4}
G.~S. Kanter, D.~Reilly, and N.~Smith, ``Practical physical-layer encryption:
  The marriage of optical noise with traditional cryptography,'' \emph{IEEE
  Commun. Mag.}, vol.~47, no.~11, pp. 74--81, 2009.

\bibitem{ple5}
M.~Jacovic, K.~Juretus, N.~Kandasamy, I.~Savidis, and K.~R. Dandekar,
  ``Physical layer encryption for wireless {OFDM} communication systems,''
  \emph{Journal of Hardware and Systems Security}, vol.~4, no.~3, pp. 230--245,
  2020.

\bibitem{ple6}
F.~Huo and G.~Gong, ``A new efficient physical layer {OFDM} encryption
  scheme,'' in \emph{IEEE INFOCOM 2014 - IEEE Conference on Computer
  Communications}, 2014, pp. 1024--1032.

\bibitem{intrinsic-interference1}
M.~Sakai, H.~Lin, and K.~Yamashita, ``Intrinsic interference based physical
  layer encryption for {OFDM}/{OQAM},'' \emph{IEEE Commun. Lett.}, vol.~21,
  no.~5, pp. 1059--1062, 2017.

\bibitem{intrinsic-interference2}
D.~Chen, J.~Lan, K.~Luo, and T.~Jiang, ``Symbol encryption and placement design
  for {OQAM}/{FMBC} systems,'' \emph{IEEE Trans. on Commun.}, vol.~70, no.~1,
  pp. 552--563, 2022.

\bibitem{noise-insertion}
R.~Ma, L.~Dai, Z.~Wang, and J.~Wang, ``Secure communication in {TDS}-{OFDM}
  system using constellation rotation and noise insertion,'' \emph{IEEE Trans.
  on Cons. Electron.}, vol.~56, no.~3, pp. 1328--1332, 2010.

\bibitem{noise-insertion2}
G.~S. Kanter, D.~Reilly, and N.~Smith, ``Practical physical-layer encryption:
  The marriage of optical noise with traditional cryptography,'' \emph{IEEE
  Commun. Mag.}, vol.~47, no.~11, pp. 74--81, 2009.

\bibitem{noise-insertion3}
D.~Reilly and G.~S. Kanter, ``Noise-enhanced encryption for physical layer
  security in an {OFDM} radio,'' in \emph{2009 IEEE Radio and Wireless
  Symposium}, 2009, pp. 344--347.

\bibitem{trung-ofdm}
J.~Zhang, A.~Marshall, R.~Woods, and T.~Q. Duong, ``Design of an {OFDM}
  physical layer encryption scheme,'' \emph{IEEE Trans. on Veh. Technol.},
  vol.~66, no.~3, pp. 2114--2127, 2017.

\bibitem{cryptographic-primitives}
W.~Li, D.~Mclernon, J.~Lei, M.~Ghogho, S.~A.~R. Zaidi, and H.~Hui,
  ``Cryptographic primitives and design frameworks of physical layer encryption
  for wireless communications,'' \emph{IEEE Access}, vol.~7, pp.
  63\,660--63\,673, 2019.

\bibitem{APLE}
W.~Li, D.~Mclernon, K.-K. Wong, S.~Wang, J.~Lei, and S.~A.~R. Zaidi,
  ``Asymmetric physical layer encryption for wireless communications,''
  \emph{IEEE Access}, vol.~7, pp. 46\,959--46\,967, 2019.

\bibitem{ple_new}
J.~Wang, G.~Han, S.~Li, F.~Zhou, and N.~Wang, ``A lightweight combined physical
  layer encryption and authentication scheme for industrial internet of
  things,'' \emph{IEEE Access}, vol.~12, pp. 6961--6970, 2024.

\bibitem{dynamic-constellation-rotation}
Y.~Hou, G.~Li, S.~Dang, L.~Hu, and A.~Hu, ``Physical layer encryption scheme
  based on dynamic constellation rotation,'' in \emph{2022 IEEE 96th Vehicular
  Technology Conference (VTC2022-Fall)}, 2022, pp. 1--5.

\bibitem{Jiang}
N.~Jiang, A.~Zhao, C.~Xue, J.~Tang, and K.~Qiu, ``Physical secure optical
  communication based on private chaotic spectral phase
  encryption/decryption,'' \emph{Opt. Lett.}, vol.~44, no.~7, pp. 1536--1539,
  Apr 2019.

\bibitem{A-Secure-Phase-Encrypted-IEEE}
A.~K. Nain, J.~Bandaru, M.~A. Zubair, and R.~Pachamuthu, ``A secure
  phase-encrypted {IEEE 802}.15.4 transceiver design,'' \emph{IEEE Trans.
  Comput.}, vol.~66, no.~8, pp. 1421--1427, 2017.

\bibitem{sotp}
X.~Hu, Z.~Wan, K.~Huang, L.~Jin, M.~Yan, Y.~Chen, and J.~Yang, ``Modulated
  symbol-based one-time pad secure transmission scheme using physical layer
  keys,'' \emph{Sci. China Inf. Sci.}, vol.~67, no.~1, p. 112303, 2023.

\bibitem{elliptic-curve-book}
J.~Silverman, \emph{The Arithmetic of Elliptic Curves}, ser. Graduate Texts in
  Mathematics.\hskip 1em plus 0.5em minus 0.4em\relax Springer New York, 2009.

\bibitem{channel_reciprocity}
J.~Zhang, R.~Woods, T.~Q. Duong, A.~Marshall, and Y.~Ding, ``Experimental study
  on channel reciprocity in wireless key generation,'' in \emph{Proc. IEEE Int.
  Workshop Signal Process. Adv. Wireless Commun. (SPAWC)}, 2016, pp. 1--5.

\bibitem{double-and-add}
D.~J. Bernstein and T.~Lange, ``Faster addition and doubling on elliptic
  curves,'' in \emph{Advances in Cryptology -- ASIACRYPT 2007}, K.~Kurosawa,
  Ed.\hskip 1em plus 0.5em minus 0.4em\relax Berlin, Heidelberg: Springer
  Berlin Heidelberg, 2007, pp. 29--50.

\bibitem{k-d-trees}
J.~L. Bentley, ``Multidimensional binary search trees used for associative
  searching,'' \emph{Commun. ACM}, vol.~18, no.~9, p. 509–517, Sep. 1975.

\bibitem{kd_tree_build}
A.~Naim, J.~Bowkett, S.~Karumanchi, P.~Tavallali, and B.~Kennedy,
  ``Deterministic iteratively built kd-tree with knn search for exact
  applications,'' 2021, available: https://arxiv.org/abs/2106.03799.

\bibitem{Elements-of-Information-Theory}
T.~Cover and J.~Thomas, \emph{Elements of Information Theory}.\hskip 1em plus
  0.5em minus 0.4em\relax Wiley, 2012.

\end{thebibliography}
\end{document}